\documentstyle[epsfig]{mn}
 
\newcommand{\cthead}[1]{\multicolumn{1}{c}{#1}}
\newcommand{\ks}{km~s$^{-1}$}
\newcommand{\kss}{km~s$^{-1}$ }
\newcommand{\etal}{et~al.}


\title[The Mopra survey] {Detection of new sources of methanol
  emission at 107 and 108~GHz with the Mopra telescope}

\author[I.E. Val'tts et al.]
{I.E. Val'tts,$^1$
 S.P. Ellingsen,$^2$
 V.I. Slysh,$^1$ 
 S.V. Kalenskii,$^1$ 
\newauthor
 R. Otrupcek$^3$ and
 M.A. Voronkov$^1$\\
  $^1$Astro Space Center of Lebedev Physical Institute, Profsoyuznaya 
84/32, 117810 Moscow, Russia \\
  $^2$School of Mathematics and Physics, University of Tasmania, GPO Box 252-21, Hobart 7001, TAS, Australia \\
  $^3$Australia Telescope National Facility, PO Box 76, Epping 2121, NSW, Australia}

\date{Received date; accepted date}
\pagerange{\pageref{firstpage}--\pageref{lastpage}}
\pubyear{1999}

\def\LaTeX{L\kern-.36em\raise.3ex\hbox{a}\kern-.15em
    T\kern-.1667em\lower.7ex\hbox{E}\kern-.125emX}

\begin{document}
\setcounter{dbltopnumber}{3}

\label{firstpage}

\maketitle

\begin{abstract}
  A southern hemisphere survey of methanol emission sources in two
  millimeter wave transitions has been carried out using the ATNF
  Mopra millimetre telescope.  Sixteen emission sources have been
  detected in the $3_1-4_0A^+$ transition of methanol at 107~GHz,
  including six new sources exhibiting class~II methanol maser
  emission features.  Combining these results with the similar
  northern hemisphere survey, a total of eleven 107-GHz methanol
  masers have been detected. A survey of the methanol emission in the
  $0_0-1_{-1}E$ transition at 108~GHz resulted in the detection of 16
  sources; one of them showing maser characteristics. This is the
  first methanol maser detected at 108~GHz, presumably of class~II.
  The results of LVG statistical equilibrium calculations confirm the
  classification of these new sources as a class~II methanol masers.
\end{abstract}

\begin{keywords}
ISM: radio lines: ISM -- masers -- surveys -- ISM: molecules
\end{keywords}

\section{Introduction}

\begin{figure}
\resizebox{1.05\hsize}{!}{\includegraphics{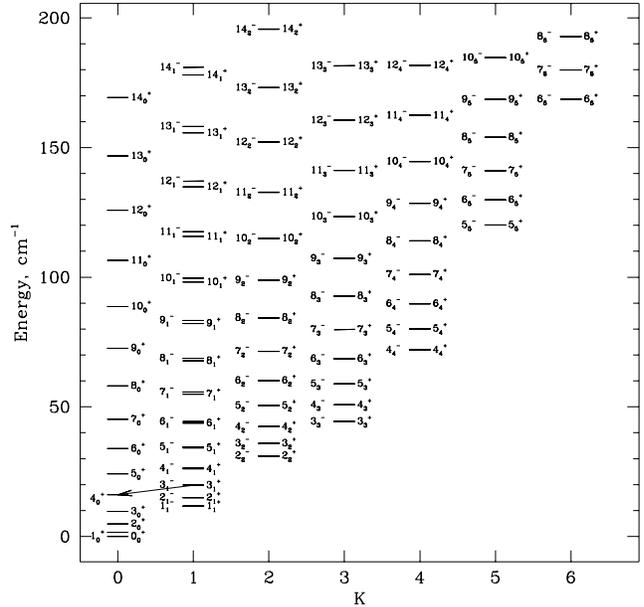}}
\caption{Energy levels for A-methanol species. The arrow represents the 107-GHz
transition}
\label{Alevels}
\end{figure}

\begin{figure}
\resizebox{1.05\hsize}{!}{\includegraphics{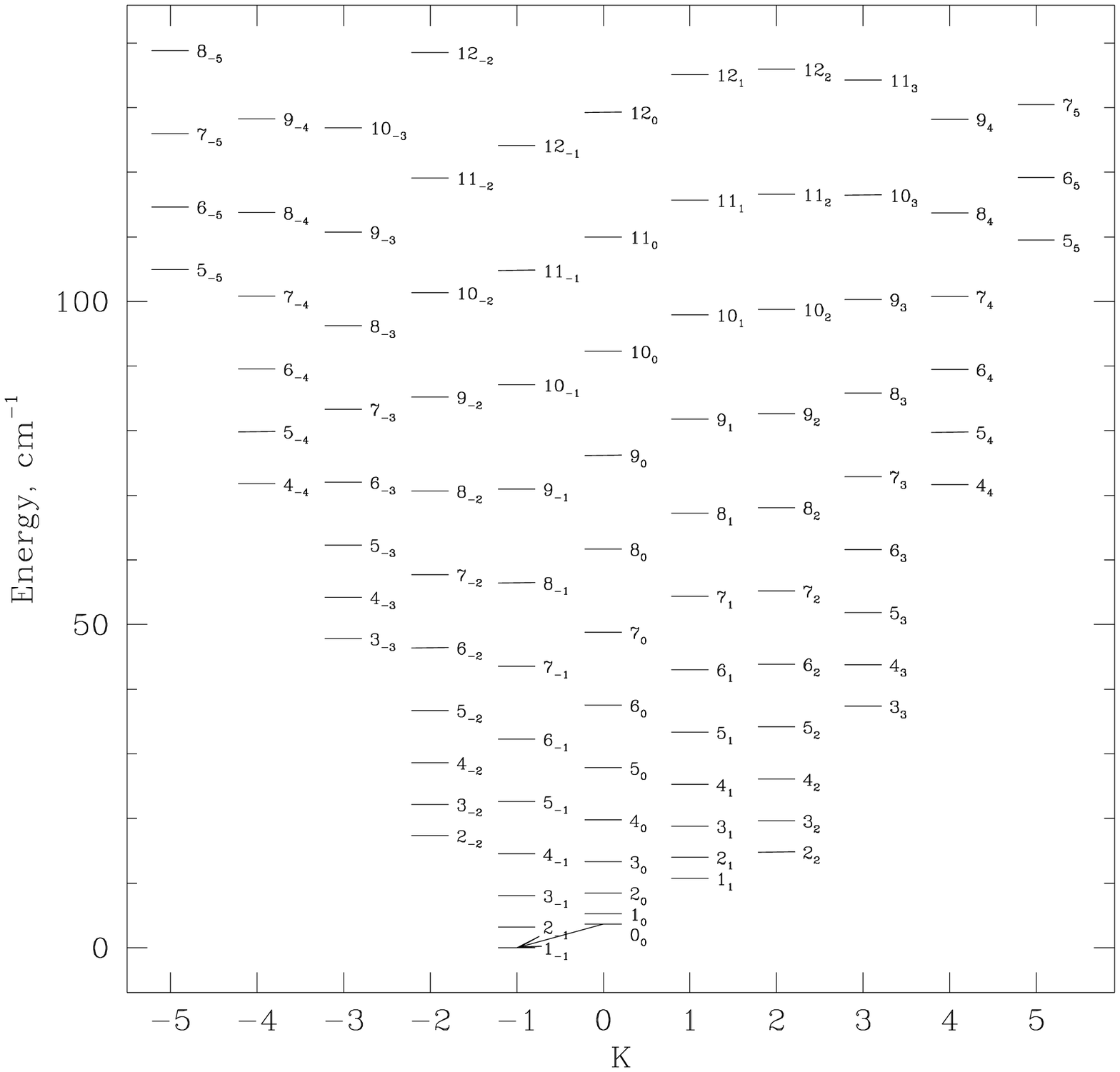}}
\caption{Energy levels for E-methanol species. The arrow represents the 108-GHz
transition}
\label{Elevels}
\end{figure}

Methanol masers are found in star forming regions, which manifest
themselves as infrared sources, molecular outflows and compact H{\sc
  ii}~regions.  They are divided into two classes, I and II
\cite{batrlaetal,MentenCl}. Class I methanol masers are not directly
associated with any known physical object, but are located in general
regions of star formation. By contrast, class~II methanol masers are
usually observed in the vicinity of newly formed luminous stars, and
are closely related to ultracompact H{\sc ii} regions and OH masers.
Both class~I and II methanol masers typically emit in several
transitions.  Figs.~\ref{Alevels} and~\ref{Elevels} show energy level
diagrams, separately for A and E-methanol. In A-methanol masers occur
mostly in transitions between K=1 and K=0 ladders, with class~I maser
transitions having their upper energy level in the K=0 ladder and
class~II maser transitions having their upper energy level in the K=1
ladder. In E-methanol most maser transitions occur between K=0 and
K=$-$1 ladders, with class~I maser transitions having their upper
energy level in the K=$-$1 ladder and class~II maser transitions
having their upper energy level in the K=0 ladder.

The strongest methanol masers are the class~II masers from the
$5_1-6_0A^+$ transition at 6.7~GHz \cite{Menten} and the $2_0-3_{-1}E$
transition at 12.2~GHz \cite{batrlaetal}.  Weaker class~II methanol
masers have been detected at millimeter wavelengths from the
$J_0-J_{-1}E$ transitions at 157~GHz (Slysh, Kalenskii \& Val'tts
1995).  Another millimeter methanol maser line from the $3_1-4_0A^+$
transition at 107~GHz (shown by the arrow in Fig.~\ref{Alevels}) was
discovered at Onsala (Sweden) \cite{v107}. Val'tts \etal\/ detected
five 107-GHz methanol maser sources, all in the northern hemisphere.
Two of them -- Cep~A and W3(OH) -- have been mapped with high angular
resolution using BIMA (Mehringer, Zhou \& Dickel 1997; Slysh \etal\/
1999a).  The masers are unresolved by BIMA, yielding a lower limit on
the brightness temperature at 107~GHz of 5$\times$10$^5$K in W3(OH).

To date the only published search for 107-GHz methanol masers is that
by Val'tts \etal\/ \shortcite{v107}.  In this paper we present the
results of an extensive search for 107-GHz methanol emission in the
southern hemisphere with the Mopra 22-m telescope (Australia).  These
observations complement the northern hemisphere survey of Val'tts
\etal\/ and completes a search for 107-GHz methanol maser emission
toward sources which show strong 6.7-GHz maser emission.  A survey of
methanol emission in the $0_0-1_{-1}E$ transition (shown by the arrow
in Fig.~\ref{Elevels}) at 108~GHz was also performed, this represents
the first search for maser emission from this transition.

\section{Observations}

The sources searched for methanol emission at 107 and 108~GHz were
selected from 6.7-GHz methanol masers detected by Menten
\shortcite{Menten}, Schutte \etal\/ \shortcite{schutte}, Caswell
\etal\/ \shortcite{casw1995a}, van der Walt \etal\/ \shortcite{walt}
and Slysh \etal\/ \shortcite{medicina}.  This included all 6.7-GHz
methanol masers with a peak flux density greater than 50~Jy south of
declination $+$21\degr.

The observations were carried out in the period from July~1 to~17,
1997, using the Mopra 22-m millimeter-wave telescope of the ATNF.  The
assumed rest frequencies of the $3_1-4_0A^+$ and $0_0-1_{-1}E$
transitions of methanol were 107.01367~GHz \cite{v107} and
108.893940~GHz \cite{lucia} respectively.  At these frequencies only
the inner 15 metres of the Mopra antenna is illuminated and the
aperture efficiency is 41\%, which implies that one Kelvin of antenna
temperature corresponds to 40~Jy for both frequencies.  The half-power
beamwidth at 107 and 108.9~GHz is 46\arcsec and 45$\farcs$5,
respectively.  The antenna pointing was checked every 12 hours through
observations of 86-GHz SiO masers, the pointing accuracy of the Mopra
antenna is 10\arcsec~rms.  The observations were performed in a
position switching mode with reference positions offset 30\arcmin.

A cryogenically cooled low-noise SIS~mixer was used in the receiver.
For the 107-GHz observations the single side-band receiver noise
temperature was 140~K and the system temperature varied between 240~K
and 325~K depending on weather conditions and the elevation of the
telescope, while at 108~GHz the single side-band receiver noise
temperature was 160~K and the system temperature varied between 255~K
and 400~K.  An ambient temperature load (assumed to have a temperature
of 290K) was regularly placed in front of the receiver to enable
calibration using the method of Kutner \& Ulich~ \shortcite{kutner},
this corrects the observed flux density for the effects of atmospheric
absorption.  Variations in the ambient temperature of a few percent
occurred during the observations, and the estimated uncertainty of the
absolute flux density scale is 10\%.

The back-end was a 64~MHz wide 1024-channel autocorrelator with a
frequency resolution of 62.5~kHz.  This yielded a velocity resolution
at 107~GHz of 0.210~\kss with uniform weighting and 0.350~\kss with
Hanning smoothing.  At 108~GHz the velocity resolution was 0.207~\kss
with uniform weighting and 0.344~\kss with Hanning smoothing.  For
each source a uniformly weighted spectrum was produced with a velocity
range of approximately 70~\kss centred on the velocity of the observed
6.7 and 12.2-GHz methanol maser, or CS thermal emission.  The spectrum
was then Hanning smoothed, as many of the sources are weak this
usually improved the signal to noise ratio.  The spectra and Gaussian
parameters (peak flux density, velocity and full width half maximum)
in all figures and tables are Hanning smoothed data unless otherwise
noted.

\begin{table*}
\begin{minipage}{105mm}
\caption{Detected 107-GHz methanol sources.  Gaussian parameters were determined from Hanning smoothed spectra, except for Orion KL, 328.81+0.63, 339.88-1.26 and 345.01+1.79}
\label{Detection107}
\begin{tabular}{lrrrrr}
\hline
\cthead{Source}&\cthead{R.A.}&\cthead{Dec.}&\cthead{Peak Flux}&\cthead{LSR radial}&\cthead{Line}\\
               &\cthead{1950}&\cthead{1950}&\cthead{Density}  &\cthead{velocity}  &\cthead{FWHM}\\
      &\cthead{(h m s)}&\cthead{(${}^\circ \quad {}\arcmin \quad { } \arcsec$)}&\cthead{(Jy)}&\cthead{(km s$^{-1}$)}&\cthead{(km s$^{-1}$)}\\      
\hline
Orion S6 &05 32 44.8&$-$05 26 00.0&8.9(0.6)&6.5(0.1)&3.9(0.3)\\
Orion KL &05 32 47.0&$-$05 24 23.0&29.3(7.3)&7.4(0.1)&1.7(0.4)\\
         && &40.0(5.5)&7.6(0.2)&5.2(0.5)\\         
S252     &06 05 54.0&$+$21 39 09.0&7.2(1.0)&10.3(0.1)&0.8(0.1)\\
305.208$+$0.206&13 08 01.6&$-$62 18 44.8&3.7(0.4)&$-$40.4(0.3)&2.5(0.7)\\
IRAS13484$-$6100&13 48 24.4&$-$61 01 30.0&4.4(0.5)&$-$55.8(0.1)&2.1(0.3)\\
323.74$-$0.26&15 27 52.0&$-$56 20 39.5&4.7(0.7)&$-$53.2(0.2)&1.6(0.3)\\
           &&&14.5(0.5)&$-$50.8(0.1)&2.4(0.1)\\
328.81$+$0.63&15 52 00.3&$-$52 34 22.2&10.6(1.4)&$-$44.1(0.1)&0.6(0.1)\\
           &&&13.1(0.5)&$-$42.1(0.1)&6.0(0.2)\\
337.403$-$0.400&16 35 07.9&$-$47 22 04.1&10.9(0.5)&$-$40.8(0.1)&4.4(0.2)\\
339.88$-$1.26&16 48 24.8&$-$46 03 33.9&58.7(4.3)&$-$38.8(0.1)&0.8(0.1)\\
           &&&49.6(1.5)&$-$37.6(0.1)&1.5(0.1)\\
           &&&29.3(5.5)&$-$36.1(0.1)&0.4(0.1)\\
           &&&19.1(2.9)&$-$35.3(0.1)&1.0(0.4)\\
           &&&67.8(2.4)&$-$34.2(0.1)&1.1(0.1)\\           
345.01$+$1.79&16 53 19.7&$-$40 09 46.0&84.5(2.4)&$-$22.3(0.1)&1.1(0.1)\\  
NGC6334F &17 17 32.4&$-$35 44 04.2&18.4(3.3)&$-$10.6(0.1)&2.5(0.4)\\
         &&&12.2(1.7)&$-$7.7(0.4)&3.5(0.6)\\    
351.78$-$0.54&17 23 20.7&$-$36 06 45.5&5.4(0.3)&$-$3.1(0.2)&8.1(0.6)\\
Sgr A$-$G&17 42 27.3&$-$29 04 35.8&9.6(1.7)&19.5(0.1)&0.4(0.2)\\
Sgr B2&17 44 10.7&$-$28 22 17.0&16.6(1.1)&57.5(0.7)&19.0(1.5)\\    
9.62$+$0.20&18 03 16.0&$-$20 31 52.0&13.9(1.1)&$-$1.0(0.1)&1.0(0.1)\\
         &&&3.8(0.4)&1.2(0.4)&6.4(0.8)\\        
W48&18 59 13.1&$+$01 09 07.0&9.2(1.3)&41.1(0.2)&1.7(0.4)\\
   &&&8.0(1.8)&44.3(0.1)&2.0(0.2)\\
\hline
\end{tabular}
\end{minipage}
\end{table*}

\section{Results}

\begin{table*}
\begin{minipage}{105mm}
\caption{Sources undetected at 107~GHz (all spectra Hanning smoothed).}
\label{Negative107}
\begin{tabular}{lrrrr}
\hline
\cthead{Source}&\cthead{R.A.}&\cthead{Dec.}&\cthead{LSR radial}&\cthead{RMS}\\
      &\cthead{1950}&\cthead{1950}&\cthead{velocity}&\\
      &\cthead{(h m s)}&\cthead{(${}^\circ \quad {}\arcmin \quad { } \arcsec$)}&\cthead{(km s$^{-1}$)}&\cthead{(Jy)}\\      
\hline
IRAS05329$-$0512&05 32 58.7&$-$05 12 11.0&11.1&1.8\\
L1641N&05 33 52.7&$-$06 54 02.0&7.2&1.5\\
Mon R2&06 05 20.0&$-$06 22 40.0&12.0&1.2\\   
S269&06 11 47.1&$+$13 50 34.0&15.0&2.1\\
IRAS07077$-$1026&07 07 43.6&$-$10 26 47.9&13.7&1.6\\
IRAS08076$-$3556&08 07 40.1&$-$35 56 07.6&5.9&1.6\\
IRAS08448$-$4343&08 44 49.3&$-$43 43 28.0&3.7&1.5\\
IRAS08470$-$4243&08 47 00.5&$-$42 43 15.0&12.3&1.6\\
IRAS09002$-$4732&09 00 12.2&$-$47 32 06.5&3.1&1.5\\
IRAS09018$-$4816&09 01 51.7&$-$48 16 41.7&10.3&1.6\\
IRAS09149$-$4743&09 07 42.2&$-$48 53 13.2&9.3&1.5\\
IRAS10184$-$5748&10 18 26.3&$-$57 48 31.5&8.9&1.6\\
IRAS10460$-$5811&10 46 03.4&$-$58 10 58.0&$-$1.8&1.7\\
305.21$+$0.21&13 08 01.7&$-$62 18 45.3&$-$38.0&1.2\\
309.921$+$0.479&13 47 11.8&$-$61 20 18.7&$-$59.6&1.4\\
IRAS14164$-$6028&14 16 24.4&$-$60 28 55.0&$-$47.8&1.7\\
318.95$-$0.20&14 57 03.8&$-$58 47 01.2&$-$34.7&1.5\\
IRAS15122$-$5801&15 12 02.6&$-$58 00 50.0&$-$62.0&1.4\\
IRAS15394$-$5358&15 39 28.4&$-$53 58 28.0&$-$38.0&1.3\\
IRAS15408$-$5356&15 40 45.8&$-$53 56 28.0&$-$43.0&1.3\\
IRAS16019$-$4903&16 02 03.2&$-$49 04 14.0&$-$23.6&1.1\\
331.28$-$0.19&16 07 38.0&$-$51 34 12.4&$-$78.0&1.2\\
IRAS16084$-$5127&16 08 24.0&$-$51 27 15.0&$-$86.0&1.4\\
331.45$-$0.18&16 08 26.2&$-$51 26 57.4&$-$88.5&1.3\\
333.129$-$0.429&16 17 13.4&$-$50 28 12.5&$-$51.2&1.9\\
333.58$-$0.02&16 17 28.0&$-$49 51 42.7&$-$35.9&1.2\\
333.45$-$0.18&16 17 32.3&$-$50 04 08.0&$-$42.5&1.2\\
333.47$-$0.17&16 17 34.7&$-$50 02 45.0&$-$42.0&1.2\\
332.96$-$0.68&16 17 35.2&$-$50 45 56.3&$-$45.9&1.3\\
334.65$-$0.02&16 25 51.1&$-$49 13 07.0&$-$30.1&1.1\\
335.79$+$0.17&16 26 04.7&$-$48 09 20.4&$-$47.5&1.1\\  
336.022$-$0.819&16 31 21.8&$-$48 40 51.0&$-$53.2&1.3\\
336.83$-$0.36&16 36 22.2&$-$47 51 48.0&$-$22.7&2.0\\
338.92$+$0.55&16 36 54.8&$-$45 36 14.0&$-$62.0&1.6\\
339.62$-$0.12&16 42 26.5&$-$45 31 18.0&$-$36.0&1.9\\
340.785$-$0.096&16 46 38.2&$-$44 37 18.6&$-$107.0&1.4\\
339.68$-$1.21&16 47 25.0&$-$46 10 59.0&$-$21.0&1.9\\
341.22$-$0.21&16 48 42.1&$-$44 21 53.0&$-$38.0&1.2\\
345.00$-$0.22&17 01 38.5&$-$41 24 59.0&$-$22.0&1.9\\
345.400$-$0.941&17 05 59.8&$-$41 32 07.0&$-$20.9&1.9\\
350.504$+$0.956&17 13 40.2&$-$36 17 54.5&$-$10.3&2.0\\
352.630$-$0.567&17 25 49.8&$-$35 25 16.3&$-$0.4&2.1\\
351.633$-$1.252&17 25 53.6&$-$36 37 50.3&$-$11.9&2.1\\
IRAS17440$-$2824&17 44 04.8&$-$28 25 49.0&49.5&1.8\\
0.64$-$0.04&17 44 08.9&$-$28 23 29.0&49.0&1.7\\
0.54$-$0.85&17 47 04.1&$-$28 54 01.0&14.8&1.7\\
IRAS17480$-$2623&17 47 44.4&$-$26 38 52.0&3.0&2.2\\
IRAS17589$-$2312&17 58 56.2&$-$23 13 53.0&26.4&1.6\\
M8E&18 01 49.7&$-$24 26 56.0&10.8&1.5\\
11.90$-$0.14&18 05 15.2&$-$18 42 21.0&43.0&1.6\\
10.30$-$0.15&18 05 57.9&$-$20 06 26.0&10.6&1.2\\
IRAS18060$-$2005&18 06 06.7&$-$20 05 34.4&11.0&1.6\\
W33B&18 10 59.3&$-$18 02 40.0&30.0&1.3\\
IRAS18128$-$1640&18 12 51.1&$-$16 39 53.0&15.0&1.5\\
IRAS18134$-$1942&18 13 28.3&$-$19 42 25.0&6.7&1.6\\
L379IRS3&18 26 32.9&$-$15 17 58.0&18.0&1.4\\
IRAS18316$-$0602&18 31 38.9&$-$06 02 23.0&41.8&1.7\\
IRAS18353$-$0628&18 35 18.0&$-$06 28 00.0&95&1.8\\
25.72$+$0.01&18 35 29.2&$-$06 27 21.4&95.7&1.8\\
\hline
\end{tabular}
\end{minipage}
\end{table*}

\begin{table*}
\begin{minipage}{105mm}
\contcaption{}
\begin{tabular}{lrrrr}
\hline
\cthead{Source}&\cthead{R.A.}&\cthead{Dec.}&\cthead{LSR radial}&\cthead{RMS}\\
      &\cthead{1950}&\cthead{1950}&\cthead{velocity}&\\
      &\cthead{(h m s)}&\cthead{(${}^\circ \quad {}\arcmin \quad { } \arcsec$)}&\cthead{(km s$^{-1}$)}&\cthead{(Jy)}\\      
\hline
W43M&18 45 37.2&$-$01 30 00.0&102.0&1.8\\
35.02$+$0.35&18 51 29.1&$-$01 57 26.0&44.0&1.7\\
IRAS18517$+$0437&18 51 48.7&$+$04 37 19.0&41.0&1.7\\
49.49$-$0.39&19 21 25.7&$+$14 24 42.0&59.0&2.0\\
\hline
\end{tabular}
\end{minipage}
\end{table*}

Emission from the 107-GHz transition of methanol was detected in 16 of
the 79 sources observed.  The detected sources are listed in
Table~\ref{Detection107} with the Gaussian parameters of spectral
features. Their spectra are shown in Fig.~\ref{Spectra107}.
Non-detections are given in Table~\ref{Negative107}.  The detection
limit after Hanning smoothing the spectra varied between 3 and 8~Jy at
the 3$-\sigma$ level.

\begin{table*}
\begin{minipage}{105mm}
\caption{Detected 108-GHz methanol sources.  Gaussian parameters were
determined from Hanning smoothed spectra, except for Orion KL, 322.16+0.64,
328.81+0.63, 337.41-0.40, 345.01+1.79, NGC6334B and 351.77-0.54} 
\label{Detection108}
\begin{tabular}{lrrrrr}
\hline
\cthead{Source}&\cthead{R.A.}&\cthead{Dec.}&\cthead{Peak Flux}&\cthead{LSR radial}&\cthead{Line}\\
               &\cthead{1950}&\cthead{1950}&\cthead{Density}  &\cthead{velocity}  &\cthead{FWHM}\\
      &\cthead{(h m s)}&\cthead{(${}^\circ \quad {}\arcmin \quad { } \arcsec$)}&\cthead{(Jy)}&\cthead{(km s$^{-1}$)}&\cthead{(km s$^{-1}$)}\\      
\hline
Orion S6 &05 32 44.8&$-$05 26 00.0&6.4(0.5)&6.5(0.1)&4.1(0.3)\\
Orion KL &05 32 47.0&$-$05 24 23.0&19.4(0.7)&7.9(0.1)&4.4(0.2)\\  
305.21$+$0.21&13 08 01.7&$-$62 18 45.3&5.7(0.5)&$-$41.1(0.2)&6.8(0.5)\\
318.95$-$0.20&14 57 03.8&$-$58 47 01.2&5.9(0.5)&$-$34.6(0.2)&3.8(0.4)\\
322.16$+$0.64&15 14 45.7&$-$56 27 28.0&12.7(0.7)&$-$57.1(0.1)&4.5(0.3)\\
328.81$+$0.63&15 52 00.3&$-$52 34 22.2&7.4(2.0)&$-$44.0(0.7)&9.4(1.0)\\
           &&&11.4(2.0)&$-$41.8(0.2)&3.9(0.6)\\
337.41$-$0.40&16 35 09.8&$-$47 22 07.0&12.1(0.6)&$-$40.8(0.1)&4.9(0.3)\\
345.01$+$1.79&16 53 19.7&$-$40 09 46.0&7.5(1.2)&$-$22.1(0.1)&1.3(0.2)\\
           &&&7.2(0.6)&$-$13.9(0.2)&4.6(0.4)\\
345.50$+$0.35&17 00 54.2&$-$40 40 18.0&3.7(0.5)&$-$18.0(0.2)&2.7(0.4)\\
345.00$-$0.22&17 01 38.5&$-$41 24 59.0&2.8(0.4)&$-$25.6(0.4)&3.9(0.8)\\
NGC6334B   &17 16 35.5&$-$35 54 44.0&11.9(0.6)&$-$7.0(0.1)&4.7(0.3)\\
NGC6334C   &17 16 54.5&$-$35 51 58.0&3.9(0.5)&$-$4.2(0.2)&3.6(0.5)\\
351.77$-$0.54&17 23 20.7&$-$36 06 45.4&11.8(0.5)&$-$3.2(0.1)&7.7(0.3)\\
9.62$+$0.20  &18 03 16.0&$-$20 31 52.9&3.4(0.5)&4.1(0.3)&2.3(0.6)\\
W33A       &18 11 43.8&$-$17 53 04.0&6.4(0.4)&36.2(0.2)&5.5(0.4)\\
49.49$-$0.39 &19 21 25.7&$+$14 54 42.0&8.9(0.4)&56.8(0.2)&8.6(0.4)\\       
\hline
\end{tabular}
\end{minipage}
\end{table*}

\begin{table*}
\begin{minipage}{105mm}
\caption{Sources undetected at 108~GHz (all spectra Hanning smoothed).}
\label{Negative108}
\begin{tabular}{lrrrr}
\hline
\cthead{Source}&\cthead{R.A.}&\cthead{Dec.}&\cthead{LSR radial}&\cthead{RMS}\\
      &\cthead{1950}&\cthead{1950}&\cthead{velocity}&\\
      &\cthead{(h m s)}&\cthead{(${}^\circ \quad {}\arcmin \quad { }\arcsec $)}&\cthead{(km s$^{-1}$)}&\cthead{(Jy)}\\      
\hline
Mon R2          &06 05 20.0&$-$06 22 40.0&12.0&1.2\\               
S252            &06 05 53.5&$+$21 39 02.0&11.0&1.6\\
192.60$-$0.05     &06 09 59.1&$+$18 00 10.0&5.0&1.5\\
S269            &06 11 47.1&$+$13 50 34.0&15.0&2.0\\
IRAS08470$-$4243&08 47 00.5&$-$42 43 15.0&12.3&1.7\\
IRAS10460$-$5811  &10 46 03.4&$-$58 10 58.0&$-$1.8&1.7\\
291.28$-$0.71     &11 09 46.7&$-$61 02 06.0&$-$30.0&1.2\\
305.20$+$0.21     &13 07 58.6&$-$62 16 42.3&$-$44.0&1.9\\
309.92$+$0.48     &13 47 11.9&$-$61 20 18.8&$-$60.0&1.3\\
IRAS14164$-$6028  &14 16 24.4&$-$60 28 55.0&$-$47.8&1.6\\
323.74$-$0.26     &15 27 52.0&$-$56 20 39.5&$-$51.0&1.8\\
IRAS15539$-$5353  &15 54 06.1&$-$53 50 47.0&$-$45.0&1.6\\
328.25$-$0.53     &15 54 07.0&$-$53 49 25.0&$-$37.0&1.9\\
336.022$-$0.819   &16 31 21.8&$-$48 40 51.0&$-$53.2&1.9\\
339.88$-$1.26   &16 48 24.8&$-$46 03 33.9&$-$39.0&1.8\\ 
354.61$+$0.47     &17 26 56.8&$-$32 41 34.0&$-$23.0&1.3\\
M8E             &18 01 49.7&$-$24 26 56.0&10.0&2.4\\    
8.68$-$0.37       &18 03 22.6&$-$21 37 24.0&43.0&1.5\\
12.89$+$0.49      &18 08 56.4&$-$17 32 14.0&39.0&1.5\\
W33B            &18 10 59.3&$-$18 02 40.0&30.0&1.7\\
23.01$+$0.41      &18 31 55.6&$-$09 03 09.0&75.0&1.7\\
25.72$+$0.01      &18 38 10.4&$-$06 24 41.0&95.7&2.4\\
29.95$-$0.02      &18 43 26.7&$-$02 42 38.0&96.0&2.2\\
35.20$-$0.74      &18 55 41.1&$+$01 36 26.0&28.0&1.7\\
W48             &18 59 13.1&$+$01 09 07.0&42.0&1.6\\
\hline
\end{tabular}
\end{minipage}
\end{table*}

\begin{figure*}
  \resizebox{\hsize}{!}{\includegraphics{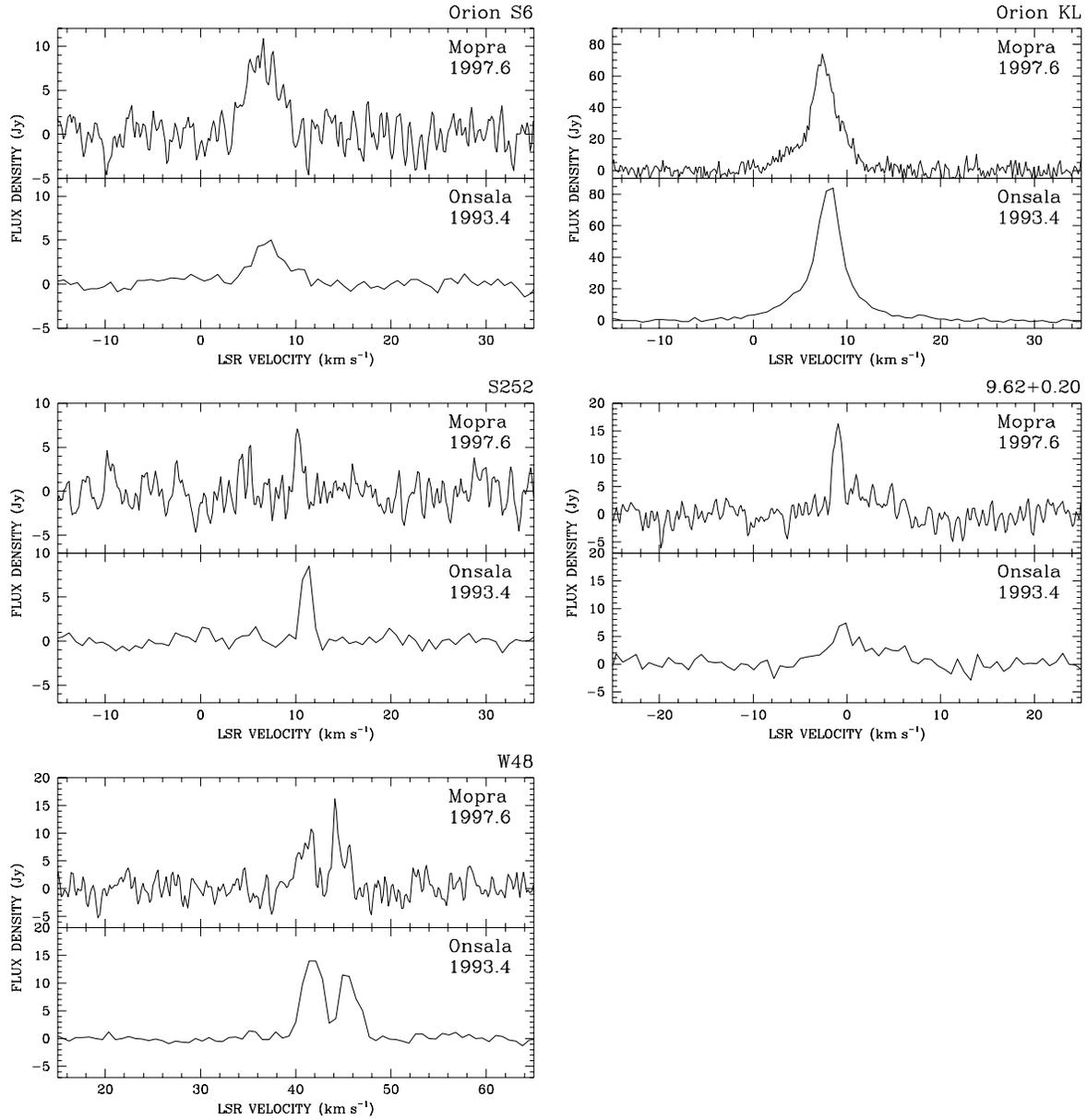}}
\caption{Comparison between Mopra and Onsala observations at 107~GHz.}
\label{MopraOnsala}
\end{figure*}

Five sources from Table~\ref{Detection107} were discovered previously
by Val'tts \etal\/ \shortcite{v107} at Onsala.  At 107~GHz there is a
rather good agreement in the radial velocity and line width of the
spectral features observed at Mopra and Onsala. The flux densities do
not agree so well.  In Orion~S6 and 9.62+0.20 the flux densities,
measured at Mopra are a factor of two larger than the Onsala flux
densities (see Fig.~\ref{MopraOnsala}). On the other hand, the S252
and W48 flux densities, measured at Mopra are a factor of 1.5 less
than those observed at Onsala. These differences are most likely due
to a combination of calibration and pointing errors and poor
signal-to-noise ratios.  For 9.62+0.20 the inconsistency may be due to
the difference in spectral resolution, or possible variability during
the time interval of 4.2 years between the two sets of observations.
Ten sources from Table~\ref{Detection107} have also been observed at
SEST by Booth \& Caswell (private communication). Some of our
undetected sources were found at SEST with better sensitivity. The
sources detected at SEST but not at Mopra are MonR2, 309.92+0.48,
338.92+0.55, 340.78-0.10, 345.00-0.22 and W33B (Booth \& Caswell,
private communication).

\begin{figure*}
\resizebox{\hsize}{!}{\includegraphics{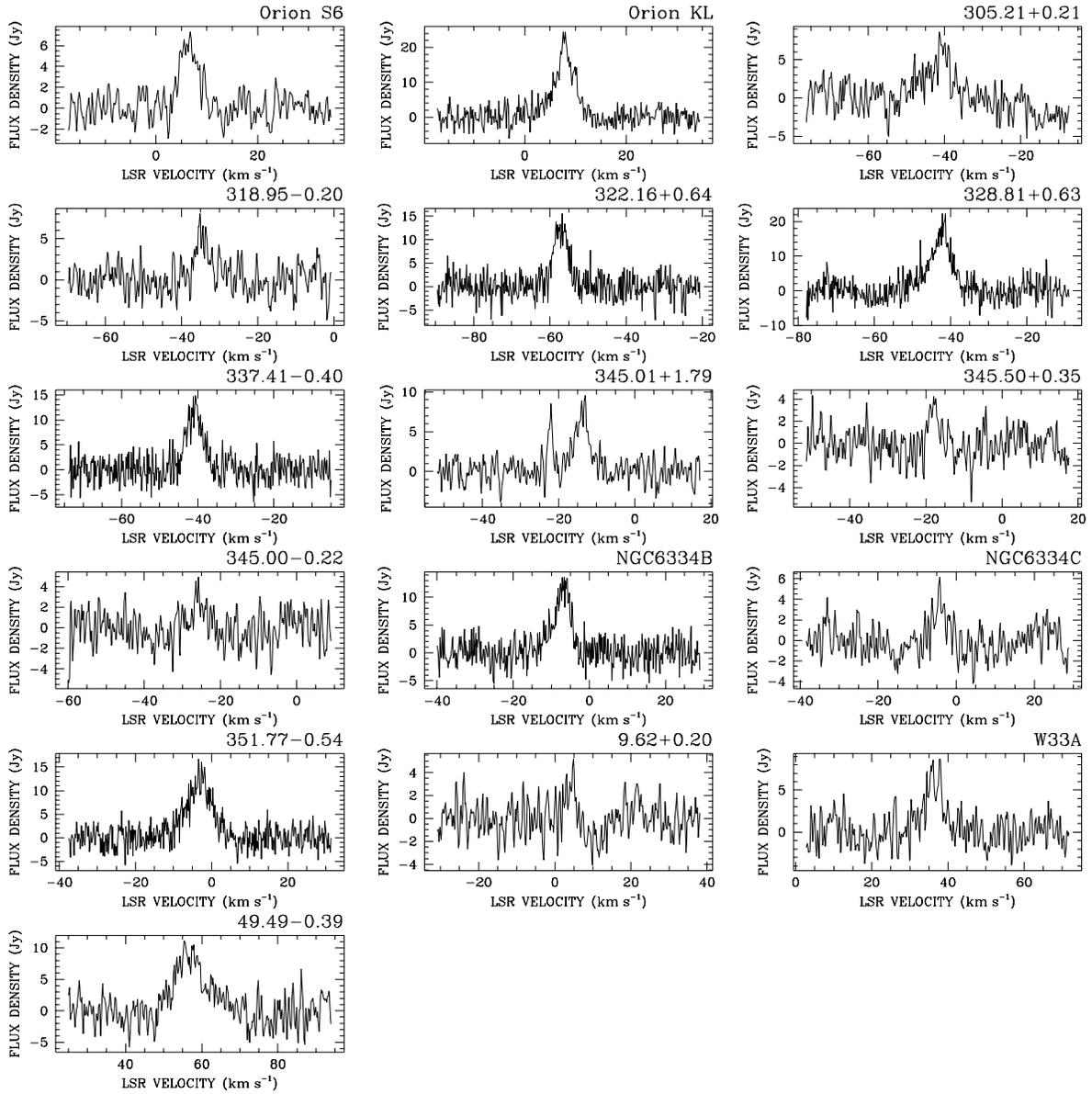}}
\vskip -10mm
\caption{108-GHz spectra.  All spectra are Hanning smoothed, except for Orion KL, 322.16+0.64, 328.81+0.63, 337.41-0.40, NGC6334B and 351.77-0.54}
\label{Spectra108}
\end{figure*}

Emission was detected toward 16 of the 41 source observed at 108~GHz.
The Gaussian parameters of the detected sources are listed in
Table~\ref{Detection108} and their spectra are shown in
Fig.~\ref{Spectra108}.  The list of non-detections is given in
Table~\ref{Negative108}. The detection limit varied between 3 and 7~Jy
at the 3$-\sigma$ level.

\begin{figure*}
\resizebox{\hsize}{!}{\includegraphics{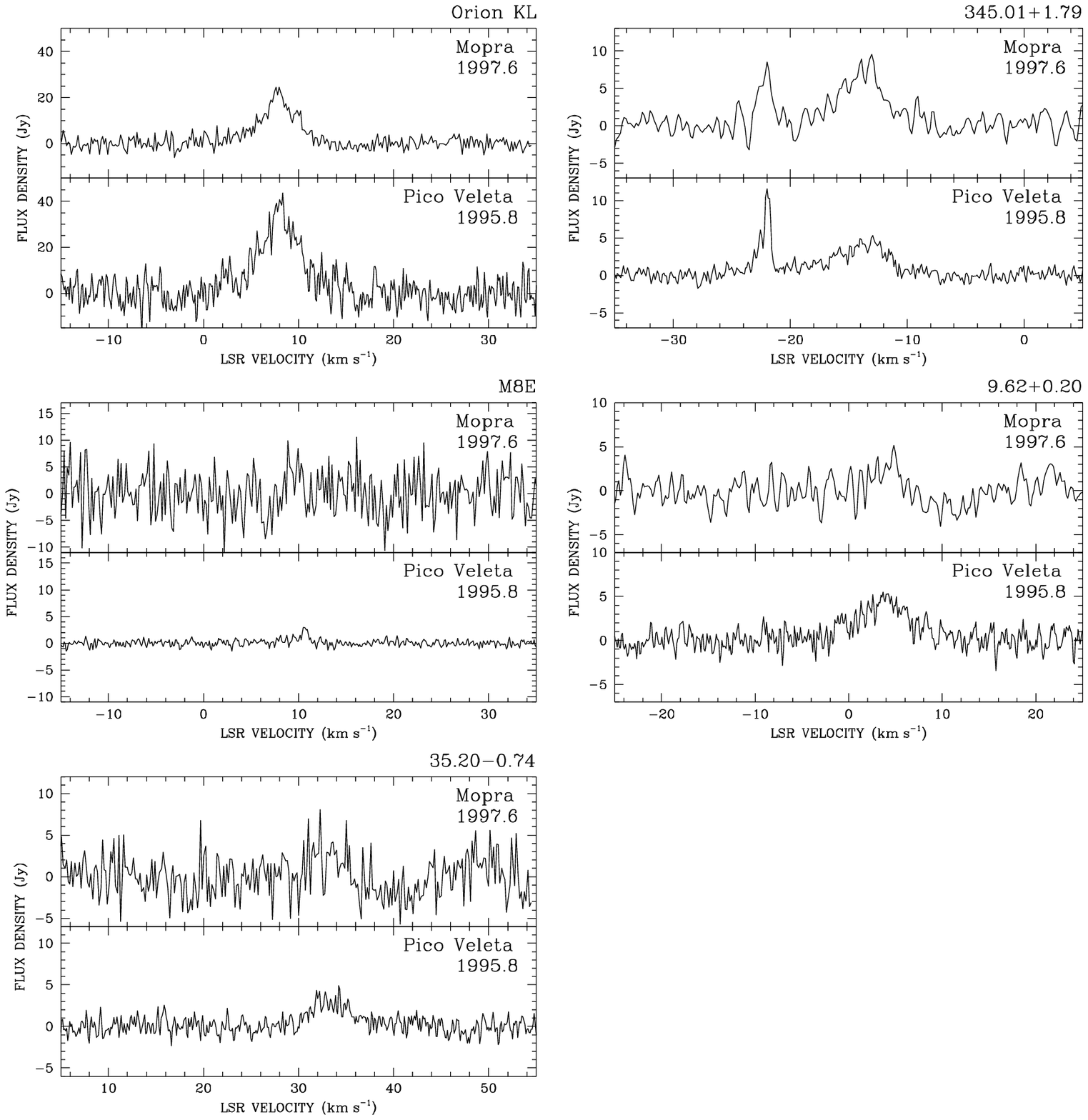}}
\caption{Comparison between Mopra and Pico Veleta observations at 108~GHz.}
\label{Mopra_PV107}
\end{figure*}

Three objects from Table~\ref{Detection108} have also been observed at
108~GHz with the 30-m telescope at Pico Veleta: Orion~KL, 345.01+1.79
and 9.62+0.20 \cite{slpv}.  Similar to the comparison of 107-GHz
sources between Mopra and Onsala observations, at 108~GHz there is
rather good agreement in the radial velocity and line width of
spectral features observed at Mopra and at Pico Veleta, but the flux
densities differ significantly.  There is a difference between
Orion~KL flux densities (see Fig.~\ref{Mopra_PV107}); which is
probably due to poor weather during the observations of Orion~KL at
Pico Veleta, which resulted in large calibration errors.  The flux
densities of the other sources are in reasonable agreement.  Four
sources from the list of those undetected at Mopra at 108~GHz
(Table~\ref{Negative108}) were also observed at Pico Veleta. Two of
them, M8E and 35.20-0.74 were detected there, however, S252 and W48
were not.  The detection of M8E and 35.20-0.74 at Pico Veleta and
non-detection in Mopra is consistent with a higher sensitivity at Pico
Veleta.

\subsection{\bf Comments on detected maser sources}

\begin{figure*}
\resizebox{\hsize}{!}{\includegraphics{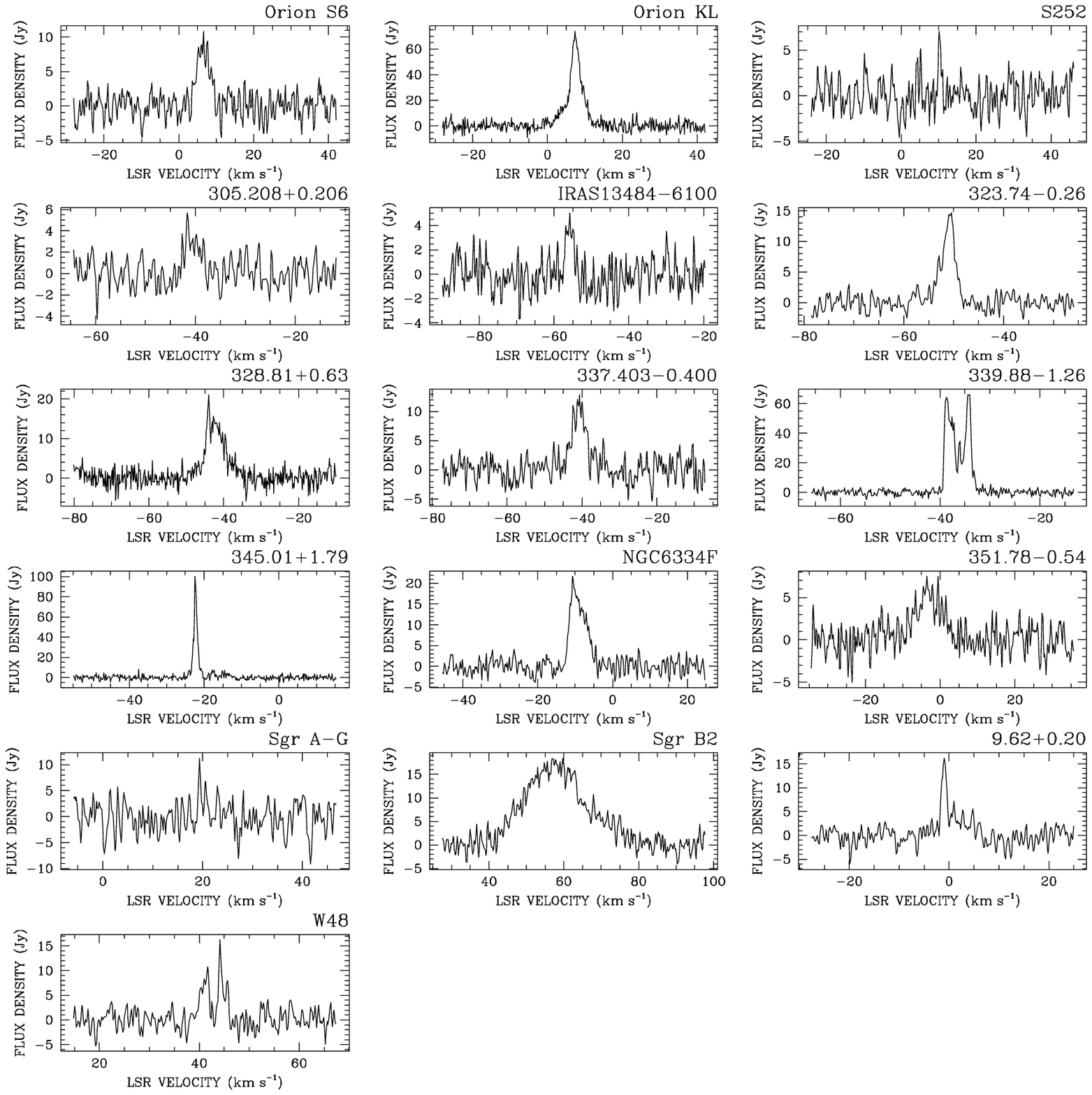}}
\vskip -10mm
\caption{107-GHz spectra.  All spectra are Hanning smoothed, except for 
  Orion KL, 328.81+0.63, 339.88-1.26 and 345.01+1.79}
\label{Spectra107}
\end{figure*}

Some of the newly detected sources show narrow lines with full width
half maximum (FWHM) of 1-2~\kss or less. This line width is less than
that of thermal lines toward these sources and suggests that the
emission is of maser origin. Another indicator of maser emission is
its anomalously high intensity; this can be seen from the comparison
of intensities at 107 and 108~GHz. In thermal emission sources the
intensity at the two frequencies is equal to within a factor of two or
three. In maser sources the difference between the intensities at 107
and 108~GHz can be very large: for example, the peak flux density of
the 107-GHz emission in 339.88$-$1.26 is at least ten times greater than
that at 108~GHz.  The overlap of several closely spaced narrow maser
emission lines can produce a spectrum with an intense wide line,
however, in such cases (e.g. 323.74$-$0.26) the intensity will be
anomalously high.  Based on these criteria we identify 9 sources in
Table~1 as 107-GHz masers: S252, 323.74$-$0.26, 328.81+0.63,
339.88$-$1.26, 345.01+1.79, NGC6334F, Sgr~A-G, 9.62+0.20, and W48.
Below we give comments on some of the maser sources.

{\bf 328.81+0.63}.  
This source shows methanol maser emission at 6.7~GHz \cite{casw1995a}
and 12.2~GHz \cite{casw1993,casw1995b}, as well
as in the class I $7_0-6_1A^+$ transition at 44.1~GHz
\cite{sl44}. Hydroxyl masers at 1.665 and 6.035~GHz are also found 
here (Caswell, Haynes \& Goss 1980; Smits 1994). The 107-GHz methanol
spectrum (Fig.~\ref{Spectra107})
shows a very narrow spike at $-$44.1~\kss and a broad component at
$-$42.1~\ks. The narrow spike with a line width 0.6~\kss has the same
radial velocity as the strongest component of 6.7-GHz maser, and must
be a 107-GHz maser.  At 108~GHz only a broad component is present.
The broad components at 107 and 108~GHz correspond to thermal CS
emission at $V_{LSR}$=$-$42.3~\kss (Bronfman, Nyman \& May 1996).

{\bf 339.88-1.26}.  
The profile of the 107-GHz emission (Fig.~\ref{Spectra107}) can be
fitted by 5 narrow components with line widths between 0.4 and
1.5~\kss (see Table~\ref{Detection107}). The velocity range of the
107-GHz emission coincides with that of the 6.7-GHz methanol maser
emission \cite{casw1995a}. Most of 107-GHz emission is clearly maser
emission, no significant thermal emission is present.  At 108~GHz the
source was not detected, although thermal CS emission is present near
the red edge of the 107-GHz emission velocity range \cite{nyman}.

{\bf 345.01+1.79}.
This is a strong methanol maser at 6.7, 12.2 and 157~GHz
\cite{casw1995a,nor1987,skv}. At both 107 and
108~GHz there is a strong narrow maser component at the same velocity
($-$22~\ks) as the 6.7, 12.2, and 157-GHz masers. This is the only
maser detected thus far at 108~GHz. A broad, probably thermal
component at 108~GHz was found at the same velocity ($-$13.9~\ks) as
thermal CS emission \cite{juv}.  For more detailed discussion of this
source see Val'tts \shortcite{v345}.

{\bf 9.62+0.20}.
This is the strongest methanol maser at 6.7~GHz \cite{Menten}.  The
107-GHz maser was discovered at Onsala by Val'tts \etal\/~\shortcite{v107}.
At Mopra we found a narrow feature at $-$1~\ks, which differs from the
radial velocity of the strongest 6.7-GHz feature (1~\ks), but there is
weak 107-GHz emission at positive radial velocities (see
Table~\ref{Detection107} and Fig.~\ref{Spectra107}).  At 108~GHz there
is only a weak broad feature at 4.1~\ks, close to the radial velocity
of thermal CS emission 5~\kss \cite{larcs}.  We have already noted
that the two spectra taken 4.2~years apart at Onsala and Mopra are
different: the narrow peak became stronger by a factor of about 3 and
its radial velocity changed by 0.5~\ks.  If this is a real
variability, it would be interesting to monitor the 107-GHz emission
from this source.

{\bf W48}.
The 107-GHz methanol maser from this source was discovered at Onsala
\cite{v107}. The spectrum consists of two moderately narrow features
as in other methanol transitions at 6.7~GHz \cite{Menten} and 157~GHz
\cite{skv}. At 108~GHz there is no detectable emission.  The thermal
CS line has a radial velocity of 43.6~\kss \cite{larcs}, roughly
coincident with the radial velocity of one of the methanol features
44.3~\ks. 

\section{Discussion}

The 107-GHz methanol masers detected in these and previous
observations have spectra which are similar to those of other class~II
methanol transitions, for example at 6.7~GHz.  Usually all the
spectral features of the 107-GHz emission have counterparts in the
6.7-GHz spectra, although the 6.7-GHz spectra typically have more
features than those at 107~GHz. The strongest 6.7-GHz features are as
a rule also the strongest features in the 107-GHz spectrum, but there
is no strong correlation between the flux density at the two
frequencies.  There are even strong 6.7-GHz masers which have not been
detected at 107~GHz, for example 309.92+0.48 has a peak flux density
of 635 Jy at 6.7~GHz but is less than 4.5~Jy~(3$-\sigma$) at 107~GHz,
and has been detected only with more sensitivity observations using
SEST (Booth and Caswell, private communication).

The $0_0-1_{-1}E$ transition at 108~GHz is also a class~II transition,
and takes place between the $K=0$ and $K=-1$ ladders, as do other
methanol-$E$ class~II maser transitions: $2_0-3_{-1}E$ and
$J_0-J_{-1}E$ at 12.2 and 157~GHz, respectively.  In comparison to the
masers at 12~GHz and 157~GHz, 108-GHz masers are extremely rare. The
only detected source is 345.01+1.79 \cite{v345}, which has a narrow
(1.3~\ks) intense maser line blue-shifted from the wide thermal line.
This has been confirmed by more sensitive Pico Veleta observations
\cite{slpv}, and the two spectra of 345.01+1.79 can be seen on
Fig.~\ref{Mopra_PV107}.  The rest of sources detected at 108~GHz show only
wide, apparently thermal emission lines.

\begin{table*}
\begin{minipage}{105mm}
\caption{Comparison of the observed flux densities and those calculated by LVG for 345.01+1.79}
\label{LVGtable}
\begin{tabular}{lrrrr}
\hline
\cthead{Transition}&\cthead{Frequency}&\cthead{Observed}
&\cthead{Calculated}&\cthead{Calculated}\\
&&\cthead{flux density}&\cthead{flux density}&\cthead{optical}\\
&\cthead{(GHz)}&\cthead{(Jy)}&\cthead{(Jy)}&\cthead{depth}\\
\hline
$5_1-6_0A^+$&6.7     & 508  & 508 &$-$8.2 \\
$2_0-3_{-1}E$&12     & 310  & 340 &$-$6.2 \\
$3_0-4_1A^+$&107     & 85.5 & 247 &$-$2.4 \\
$0_0-1_{-1}E$&108    &  9.5 & 109 &$-$1.6 \\
$4_0-4_{-1}E$&157    & 54.9 & 136 &$-$1.6 \\
$2_1-3_0A^+$&157     & 21.4 & 112 &$-$1.4 \\  
\hline
\end{tabular}
\end{minipage}
\end{table*}

It is well established that class~II methanol masers are typically
associated with ultracompact H{\sc ii}~regions \cite{phil98,walsh}.
The interpretation that the 107 and 108-GHz emission is produced by
masing is supported by the results of statistical equilibrium
calculations of methanol level excitation by external
submillimeter/far~infrared radiation from an ultracompact H{\sc
  ii}~region, and de-excitation by collisions with cold gas particles.
In Table~\ref{LVGtable} we present calculations of the relative
intensity and optical depth of several class~II methanol transitions.
The observed and calculated flux densities in the strongest transition
$5_1-6_0A^+$ at 6.7~GHz were made equal. The modelling was performed
with LVG code kindly supplied to us by Dr. Walmsley.  The assumed
molecular gas density was 10$^7$~cm$^{-3}$, temperature 50~K and
methanol density divided by velocity gradient $0.5\times
10^{-1}$~cm$^{-3}$~(km~s$^{-1}$~pc$^{-1}$)$^{-1}$.  The source was
illuminated by a compact background H{\sc ii}~region with an emission
measure of 10$^{12}$~cm$^{-6}$pc. The dilution factor was set to 0.2.
The maser is excited by the combined action of the radiation from the
compact H{\sc ii}~region and collisions between methanol and hydrogen
molecules. The collision selection rules employed in the model are
based on the paper by Lees \& Haque~\shortcite{leehaq} and imply that
the $\Delta K=0$ collisions are preferred.

All the transitions from Table~\ref{LVGtable} are inverted in our
model.  One can see in Table~\ref{LVGtable} a qualitative agreement
between observed and calculated flux densities for all transitions;
however, the quantitative agreement is good for the $2_0-3_{-1}E$
transition at 12~GHz, but at higher frequencies the calculated flux
density is systematically higher by a factor 3 to 10. A more
complicated model with regions of different temperature and density
may well give a better agreement with observations.  In addition, the
required emission measure $EM=10^{12}$~cm$^{-6}$pc seems to be too
high, as no compact H{\sc ii}~region with an emission measure of
greater than 5$\times$10$^{10}$~cm$^{-6}$pc has been observed to date.
Nevertheless, the qualitative agreement of the theory and observations
for the $3_1-4_0A^+$ and $0_0-1_{-1}E$ transitions implies that both
of them are class~II methanol masers and that the maser emission in
these transitions is excited by external radiation.

\section{Summary}
\begin{enumerate}
\renewcommand{\theenumi}{\arabic{enumi}.}
\item As a result of a survey in the southern hemisphere 16 methanol
  emission sources were detected in the $3_1-4_0A^+$ transition at
  107~GHz. This survey together with a similar survey made with the
  Onsala telescope completes a whole sky survey of methanol emission
  at 107~GHz.
\item Six new 107-GHz methanol masers were detected among 16
  emission sources.  Together with masers from the northern survey
  this makes a total of eleven 107-GHz methanol masers. They belong to
  class~II.
\item A survey for methanol emission in the $0_0-1_{-1}E$ transition
  at 108~GHz has been carried out in the southern hemisphere with the
  detection of 16 new emission sources. One maser was found, also of
  class~II.
\item The relative intensity of the class~II methanol transitions is
  consistent with a maser model with radiative excitation and
  collisional de-excitation.
\end{enumerate}
\section{Acknowledgements}

Authors are grateful to Dr.~R.~Booth and Dr.~J.~Caswell for
communicating unpublished results from SEST observations at 107~GHz.
I.E.V. is grateful to the ATNF for the hospitality, and to the staff
of Mopra observatory for the help with the observations.  The
Australia Telescope is funded by the Commonwealth of Australia for
operation as a National Facility managed by CSIRO.  Travel to
Australia for I.E.V. was aided by grant 96/1990 from the Australian
Department of Industry, Science and Tourism.  The work of I.E.V.,
V.I.S., S.V.K. and M.A.V. was partly supported by the grants
95-02-05826 and 98-02-16916 from the Russian Foundation for Basic
Research.  S.P.E thanks the Queen's trust for the computing system
used to process the data from these observations.

\label{lastpage}

\begin{thebibliography}{}
  
\bibitem[\protect\citename{Batrla et al.\ }1987]{batrlaetal} Batrla
  W., Matthews H.E., Menten K.M., Walmsley C.M., 1987, Nat., 326, 49

\bibitem[\protect\citename{Bronfman et al.\ }1996]{nyman} Bronfman L.,
  Nyman L.A., May J., 1996, A\&AS, 115, 81

\bibitem[\protect\citename{Caswell et al.\ }1980]{casw1980} Caswell
  J.L., Haynes R.F., Goss W.M., 1980, Aust. J. Phys., 33, 639

\bibitem[\protect\citename{Caswell et al.\ }1993]{casw1993} Caswell
  J.L., Gardner F.F., Norris R.P., Wellington K.J., McCutcheon W.H.,
  Peng R.S., 1993, MNRAS, 260, 425 
  
\bibitem[\protect\citename{Caswell et al.\ }1995a]{casw1995a} Caswell
  J.L., Vaile, R.A., Ellingsen, S.P., Whiteoak, J.B., Norris, R.P.,
  1995a, MNRAS, 272, 96
  
\bibitem[\protect\citename{Caswell et al.\ }1995b]{casw1995b} Caswell
  J.L., Vaile R.A., Ellingsen S.P., Norris R.P., 1995b, MNRAS, 274,
  1126

\bibitem[\protect\citename{De Lucia et al.\ }1989]{lucia} De
  Lucia F.C., Herbst E., Anderson T., Helminger P., 1989, J. Mol.
  Spectr., 134, 395 
  
\bibitem[\protect\citename{Juvella }1996]{juv} Juvella M., 1996,
  A\&AS, 118, 191
  
\bibitem[\protect\citename{Kutner \& Ulich }1981]{kutner} Kutner,
  M.L., Ulich, B.L., 1981, ApJ, 250, 341
  
\bibitem[\protect\citename{Larionov et al.\ }1999]{larcs} Larionov
  G.M., Val'tts I.E., Winnberg A., Johansson L.E.B., Booth R.S.,
  Golubev V.V., 1999, A\&AS, submitted

\bibitem[\protect\citename{Lees \& Haque }1974]{leehaq} Lees R.M.,
  Haque S.S., 1974, Can. Journ. of Phys., 52, 2250
  
\bibitem[\protect\citename{Mehringer et al.\ }1997]{mehr} Mehringer
  D.M., Zhou S., Dickel H.R., 1997, ApJ, 475, L57
  
\bibitem[\protect\citename{Menten }1991a]{MentenCl} Menten K.M., 1991,
  In: Haschick A.D., Ho P.T.P.(eds.) Proc. Third Haystack Observatory
  Meeting, MA, USA, Skylines, 119
  
\bibitem[\protect\citename{Menten }1991b]{Menten} Menten K.M., 1991b,
  ApJ, 380, L75
  
\bibitem[\protect\citename{Norris et al.\ }1987]{nor1987} Norris R.P.,
  Caswell J.L., Gardner F.F., Wellington K.J., 1987, ApJ, 321, L159
  
  
\bibitem[\protect\citename{Phillips et al.\ }1998]{phil98} Phillips
  C.J., Norris R.P., Ellingsen S.P., McCulloch P.M., 1998, MNRAS, 300,
  1131
  
\bibitem[\protect\citename{Schutte et al.\ }1993]{schutte} Schutte
  A.J., Walt D.J. van der, Gaylard M.J., MacLeod G.C., 1993, MNRAS,
  261, 783
  
\bibitem[\protect\citename{Slysh et al.\ }1994]{sl44} Slysh V.I.,
  Kalenskii S.V., Val'tts I.E., Otrupcek R., 1994, MNRAS, 268, 464
  
\bibitem[\protect\citename{Slysh et al.\ }1995]{skv} Slysh V.I.,
  Kalenskii S.V., Val'tts I.E., 1995, ApJ, 442, 668
  
\bibitem[\protect\citename{Slysh et al.\ }1999a]{slw3oh} Slysh V.I.,
  Val'tts I.E., Kalenskii S.V., Larionov G.M., 1999a, Astron. Reports,
  in press
  
\bibitem[\protect\citename{Slysh et al.\ }1999b]{medicina} Slysh V.I.,
  Val'tts I.E., Kalenskii S.V., Voronkov M.A., Palagi F., Tofani G.,
  Catarzi M., 1999b, A\&AS, 134, 115
  
\bibitem[\protect\citename{Slysh et al.\ }1999c]{slpv} Slysh V.I.,
  Val'tts I.E., Kalenskii S.V., Voronkov M.A., 1999, in preparation
  
\bibitem[\protect\citename{Smits }1994]{smits} Smits D.P., 1994,
  MNRAS, 269, 11p
  
\bibitem[\protect\citename{Val'tts }1998]{v345} Val'tts, 1998, Astron.
  Letters, 24, 788
  
\bibitem[\protect\citename{Val'tts et al.\ }1995]{v107} Val'tts I.E.,
  Dzura A.M., Kalenskii S.V., Slysh V.I., Booth R.S., Winnberg A.,
  1995, A\&A, 294, 825
  
\bibitem[\protect\citename{Walsh et al.\ }1997]{walsh} Walsh A.J.,
  Hyland A.R., Robinson G., Burton, M.G., 1997, MNRAS, 291, 261
  
\bibitem[\protect\citename{van der Walt et al.\ }1995]{walt}van der
  Walt D.J., Gaylard M.J., MacLeod G.C., 1995, A\&AS, 110, 81

\end{thebibliography}
\end{document}